\documentclass[aps,prb,twocolumn,showpacs,floatfix,superscriptaddress,amssymb,amsmath]{revtex4}

\usepackage{graphicx}
\usepackage{dcolumn}
\usepackage{bm}

\begin{document}

\title{Spin polarization control via magnetic barriers and spin-orbit effects}

\author{Anh T. Ngo}
\affiliation{Department of Physics and Astronomy, and Nanoscale
and Quantum Phenomena Institute, Ohio University, Athens, Ohio
45701-2979}
\author{J. M. Villas-B\^{o}as }
\affiliation{Department of Physics and Astronomy, and Nanoscale
and Quantum Phenomena Institute, Ohio University, Athens, Ohio
45701-2979}
\affiliation{Walter Schottky Institut and Physik Department, Technische Universit\"at  M\"unchen, Am Coulombwall 3, D-85748 Garching, Germany}
\author{Sergio E. Ulloa}
\affiliation{Department of Physics and Astronomy, and Nanoscale
and Quantum Phenomena Institute, Ohio University, Athens, Ohio
45701-2979}

\date{5 June 2008}

\begin{abstract}
We investigate the spin-dependent
transport properties of two-dimensional electron gas (2DEG)
 systems formed in diluted magnetic semiconductors and
 in the presence of Rashba spin-orbit interaction in the framework of the scattering matrix approach.
  We focus on nanostructures consisting of realistic magnetic barriers produced
   by the deposition of ferromagnetic strips on heterostructures.
  We calculate spin-dependente conductance of such barrier systems and show that the magnetization pattern
  of the strips, the tunable spin-orbit coupling,
   and the enhanced Zeeman splitting have a strong effect on the conductance of the structure.
    We describe how these effects can be employed in the efficient control of spin polarization
     via the application of moderate fields.

\end{abstract}


\maketitle


Spin-orbit coupling in semiconductors intrinsically connects the spin
of an electron to its momentum, \cite{0} providing a pathway for electrically
initializing and manipulating electron spins for applications in spintronics \cite{1,1b}
 and spin-based quantum information processing. \cite{2}
 This coupling can be regulated with quantum confinement in semiconductor heterostructures through band
  structure engineering, as well as
   by the application of external electric fields,
   as in the celebrated spin field-effect transistor proposed by Datta and Das. \cite{DD}
Using diluted magnetic semiconductors (DMS)
in such systems provides an additional degree of control of the transport properties.
 In particular, when an external magnetic field is applied, the magnetic dopant spins align,
 giving rise to a strong exchange field that acts on the electron spin.
 This $s$-$d$ exchange interaction between the electron spin in the conduction band and the
  localized magnetic ions induces a giant Zeeman splitting.

In this communication we investigate the spin-dependent transport properties of two-dimensional electron
gas (2DEG) systems formed in diluted magnetic semiconductors
 and take into account the electric-field--dependent Rashba spin-orbit
  (SOI) interaction. We focus our attention on nanostructures consisting of realistic
   magnetic barriers produced by the deposition of ferromagnetic strips near the heterostructures, \cite{3}
   providing a relatively strong inhomogeneous magnetic field on the 2DEG.\@
We show how the conductance of the 2DEG depends strongly on the magnetization pattern of the strips,
as well as on geometry and externally applied electric fields.
 We demonstrate that significant spin polarization (exceeding 50\%) can be obtained at
 low temperatures for ferromagnetic strips of typical dimensions and magnetization.

\begin{figure}[tbh]
\begin{minipage}[t]{\linewidth}
\includegraphics[width=0.8\linewidth]{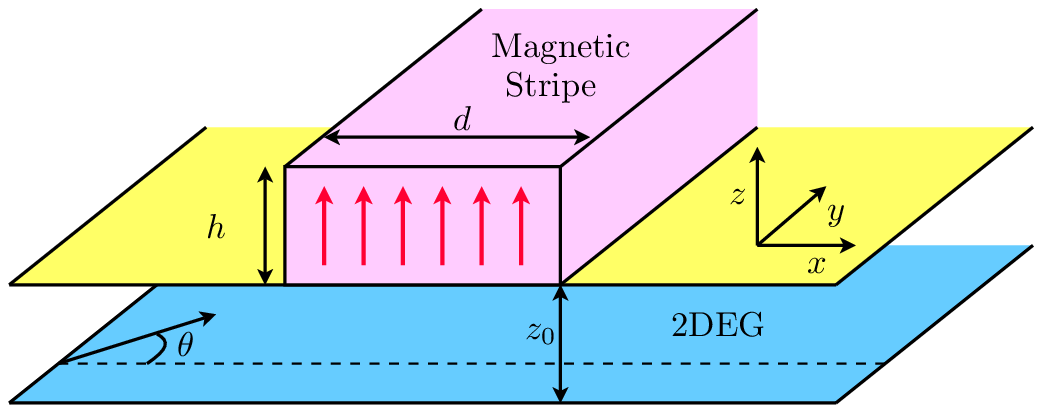}
\end{minipage}\vspace{0.1cm}
\begin{minipage}[t]{\linewidth}
\includegraphics[width=1\linewidth]{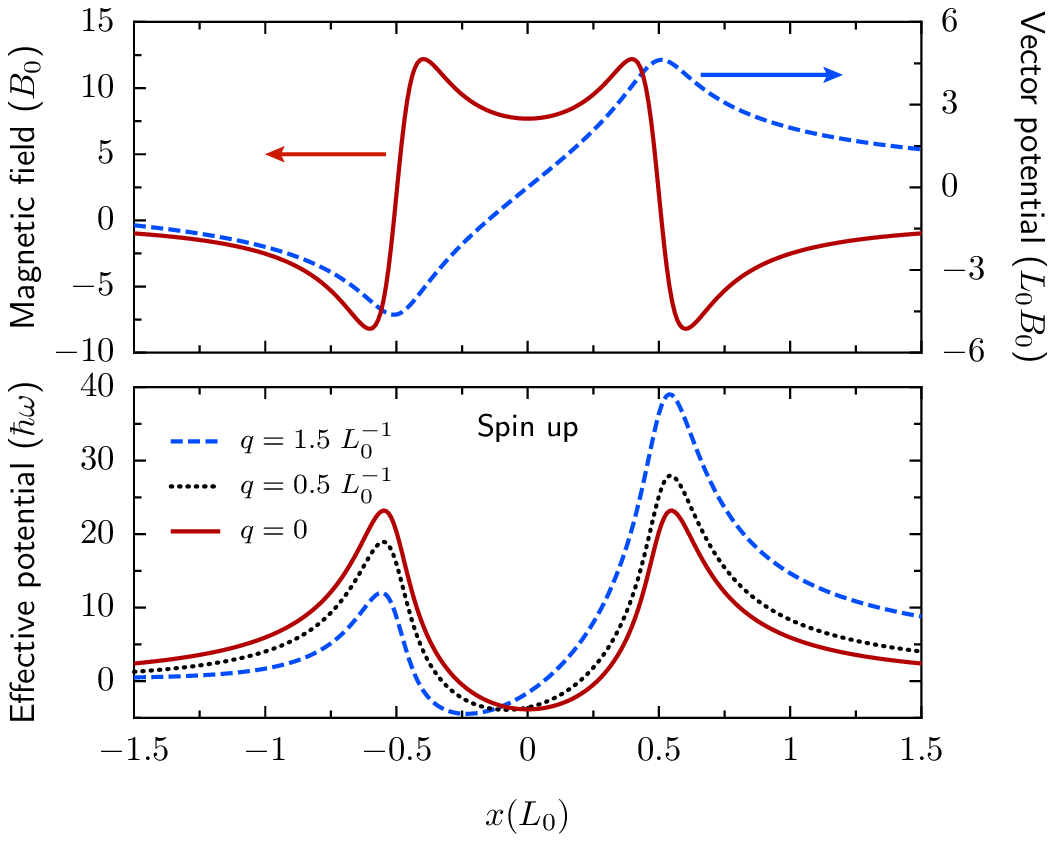}
\end{minipage}
\caption{(color online) Top diagram illustrates magnetic strip with perpendicular magnetization over a 2DEG layer. The magnetization of the strip results in the magnetic barrier and vector potential shown in middle panel.  Bottom panel shows effective potential $V_e$ for different $q$ values (Eq.\ \ref{Ve}).  Structural parameters chosen as $d=1$ and $z_{0}=0.1$, in units of the magnetic length $L_0=\sqrt{\hbar/eB_0} = 8.1$ nm for $B_0=0.1$T.} \label{fig1}
\end{figure}

Figure \ref{fig1} illustrates the type of magnetic barrier which can be created by the deposition of a ferromagnetic strip on the surface of a heterostructures.
In this case the magnetization is assumed perpendicular to the
2DEG located at a distance $z_{0}$ below the surface.
 For this strip of width $d$ and thickness $h$, the magnetic field along the $\hat{z}$-axis is given by, \cite{3}
\begin{equation}
B =B_{0}[f(x+d/2)-f(x-d/2)] \, ,
\end{equation}
where $B_{0}=M_{0}h/d$,
$f(x)=2xd/(x^{2}+z_{0}^{2})$, and $M_{0}$ is the magnetization of the
ferromagnetic strip.  The in-plane field component can be assumed negligible for this configuration.\cite{3}
We should mention that deposition of dysprosium strips result in magnetic fields estimated to be as large as
1T in the vicinity of the 2DEG; \cite{Weiss}  from this we take $B_0=0.1$T to be a realistic value.


The Hamiltonian of the 2DEG (in the $xy$ plane) is realized in the lowest subband
of the semiconductor heterostructure with effective mass $m^{*}$, and includes a
Rashba SOI due to $z$ confinement characterized by the coupling constant
 $\alpha_{z}$, Zeeman term with electronic $g$ factor, and the $s$-$d$ exchange interaction:

\begin{eqnarray}
H&&=\frac{1}{2m^{*}}(\Pi_{x}^{2}+\Pi_{y}^{2})-\frac{\alpha_{z}}{\hbar}(\sigma_{x}\Pi_{y}-\sigma_{y}\Pi_{x})
\nonumber\\
&&+\frac{g\mu_{B}\sigma_{z}}{2}B(x)+J_{sd}\sum_{i}s(r) \cdot
S(R_{i})\delta(r-R_{i}), \label{hamiltonian}
\end{eqnarray}
where $\Pi_{\mu}$ denotes the kinetic momentum, and $\sigma_{\mu}$ the Pauli spin matrices.
 $S$ is the spin of the localized $3 d^{5}$ electrons of Mn$^{2+}$ ions with $S=5/2$,
 and $s$ is the electron spin in the 2DEG.\@ We assume that the magnetic ions are distributed homogeneously
 in the DMS, and that the extended nature of the electronic wave function
 spans a large number of magnetic ions, allowing the use of a molecular-field approximation to replace the magnetic-ion spin operator $S_{i}$ with the thermal and spatial average $\langle S_{z}\rangle$ along the external magnetic field direction. This approach has been proven to be suitable in previous studies. \cite{5,6}
Correspondingly, the exchange interaction in Eq.\ (\ref{hamiltonian})
 can be written as a Zeeman-like term, $J_{sd}\langle S_{z}\rangle \sigma_{z}$,
 where $\langle S_{z} \rangle=(5/2)B_{J}(Sg\mu_{B}B/k_{B}(T+T_{0}))$, with $B_{J}(x)$
  as the Brillouin function, $T_{0}$ accounts for the reduced single-ion contribution due
  to the antiferromagnetic Mn-Mn coupling, and $k_{B}$ is the Boltzmann constant.
   The parameters used in the calculation are
    $J_{sd}= -0.22x$ eV, $x=0.014$, $T_{0}=40$K, \cite{7}
     and $T=1$K.   For convenience we express all quantities in dimensionless units,
  using $\omega=eB_{0}/m^{*}$,  and magnetic length $L_{0}=\sqrt{\hbar/eB_{0}}$.
   For Ga$_{1-x}$Mn$_{x}$As, $m^{*}=0.067 m_{e}$,
    one gets in $L_{0}= 8.1$ nm and $\hbar\omega=0.17$ meV, for $B_0=0.1$ T as above.

The two-dimensional Schr\"{o}dinger equation $H\Psi=E\Psi$
 has solutions of the form $\Psi(x,y)=e^{iqy}\psi(x)$,
  where $E$ is the total energy of the electron, and $q$
  is the electron wave vector in $y$ direction.  $\psi(x)$ satisfies
\begin{eqnarray}
\left \{\frac{d^{2}}{dx^{2}}-
V_e(x,\sigma_z) + \alpha_{z}^{*}[\sigma_{x}(q+A(x)) \right. \\ \nonumber
\left.  - \sigma_{y}(-i\frac{d}{dx})] \right \}\psi(x)
=2E\psi(x) \label{eq3} \\
 V_e(x,\sigma_z)=[q+A(x)]^{2} -[J_{sd}^{*}\langle S_{z}\rangle
+\frac{1}{2}g^{*}B(x) ]\sigma_{z} \label{Ve}
\end{eqnarray}
where $\alpha^{*}_{z}=2\alpha_{z}/L_{0}\hbar\omega$,
$J_{sd}^{*}=2J_{sd}/\hbar\omega$, $g^*=g \mu_B B_0 / \hbar \omega$, and $V_e$ is the effective dynamical potential.  We use the Landau-gauge, $\vec{A}(x)=(0,A(x),0)$, and $B(x)=dA(x)/dx$.

For a magnetic barrier structure as that shown in Fig.\ \ref{fig1},
 it is difficult to solve Eq.\ (\ref{eq3}) analytically.
   However, a numerical approach which divides the region of the
   barrier $[x_{i},x_{f}]$ into $N(\gg1)$ segments of width $L=(x_{f}-x_{i})/N$ is possible.
    \cite{4} For small $L$ (large $N$) the vector potential in each of the segments can be
     treated as constant, so that in each $j$th segment
      one has solutions given by $\psi_{j}(x)_{\pm}=e^{ik_{j}x}\chi_{\pm}(k_{j})$, with spinors
\begin{eqnarray}
\chi_{\pm}(k_{j})&=&\binom{\chi_{\pm}^{1}}{\chi_{\pm}^{2}}\ ,
\label{chieq}
\end{eqnarray}
where
\begin{eqnarray}
\chi_{\pm}^{1}&=&\frac{-\alpha_{z}^{*}[q+A(x_j)+ik_{j}]}{
\sqrt{2\lambda_{\pm}(\lambda_{\pm}-Z_j)}} \, ,
\nonumber\\
\chi_{\pm}^{2}&=&\frac{\lambda_{\pm}-Z_j}{
\sqrt{2\lambda_{\pm}(\lambda_{\pm}-Z_j)}} \, , \nonumber \\
Z_j &=& J_{sd}^{*} \langle S_{z}\rangle+\frac{1}{2}g^{*}B(x_j) \, ,
\end{eqnarray}
and
\begin{equation}
\lambda_{\pm}=\pm \left
\{Z_j^{2}+|\alpha_{z}^{*}[q+A(x_j)+ik_{j}]|^{2}\right \}^{1/2},
\end{equation}
where  $k_{j}$ is the solution of the equation
$[k_{j}^{2}+(q+A(x_j))^{2}-2E]^2= \lambda_{\pm}^2$. We consider the
transmission of electrons with initial wave vectors $k^{+}$ and
$k^{-}$ corresponding to the region $x \ll x_{i}$, which are the
solutions of the problem with $J_{sd}\langle S_z \rangle=0$, and
$\alpha_z=0$ away from the magnetic barrier region.
  It is clear that the transmission probabilities for electrons of spin eigenstates
   ``$+$" and ``$-$" in (\ref{chieq}) are different due to the SOI,
   as well as by the enhanced and inhomogeneous Zeeman splitting in (\ref{hamiltonian}).
     Since the wave vector in the $y$ direction is fixed,
     on both incident and outgoing regions of the magnetic barrier, the wave function can
     be written as
      $\psi(x)=e^{ik^{+}x}\chi_{+}(k^{+})+r^{+}e^{-ik^{+}x}\chi_{+}(-k^{+})+e^{ik^{-}x}\chi_{-}(k^{-})+r^{-}e^{-ik^{-}x}\chi_{-}(-k^{-})$ for $x<x_{i}$; and as $t^{+}e^{ik^{+}x}\chi_{+}(k^{+})+t^{-}e^{ik^{-}x}\chi_{-}(k^{-})$ for $x>x_{f}$. Here $t^{\pm}$ and $r^{\pm}$ are
       the transmission and reflection coefficients for each spin state $\chi_{\pm}$. Taking into account the boundary conditions,
       which require that the wave function be continuous at the interfaces, a system of linear equations for $t^{\pm}$ and
       $r^{\pm}$ can be derived.  Once $t^{\pm}$ and $r^{\pm}$ are known,
      it is straightforward to obtain the spin dependent transmission coefficients $T_{ss'}(E,q)$, where $s'$ ($s$)
       is the incident (outgoing) spin, with $s,s'=\uparrow$ or $\downarrow$.
       Note that for incident electrons with spin $s'=\uparrow$
      there are two kinds of transmission coefficients, $T_{\uparrow\uparrow}$, and $T_{\downarrow\uparrow}$,
       as the SOI produces precession of the electron spin as it propagates.
       A similar effect occurs for incident electrons with spin $s'=\downarrow$, although the presence of the
       magnetic field breaks time-reversal invariance which results in a spin-filtering effect, as we will see below.

In the ballistic regime the spin-dependent conductance can be calculated from the Landau-B\"{u}ttiker formula, \cite{3,4}
\begin{equation}
G_{ss'}(T)=\int_{0}^{\infty} g_{ss'}(E)\left(-\frac{\partial
f(E,T)}{\partial T}\right) \, dE \, ,
\end{equation}
where
\begin{equation}
g_{ss'}(T)=G_{0}\int_{-\pi/2}^{\pi/2}
T_{ss'}(E,\sqrt{2E}\sin(\theta))\cos(\theta) \, d\theta \, ,
\end{equation}
$\theta$ is the angle of incidence with respect to the $x$-direction
(see top panel Fig.\ \ref{fig1}), $f(E,T)$ is the Fermi-Dirac distribution function,
 and $G_{0}= e^2m^*v_FL_y/\hbar^2=(e^2/\hbar) k_F L_y$, where $L_y$ is the length
  of the structure in $y$ direction and $v_F=\hbar k_F$ is the Fermi velocity.

\begin{figure}[tb]
\includegraphics[width=1\linewidth]{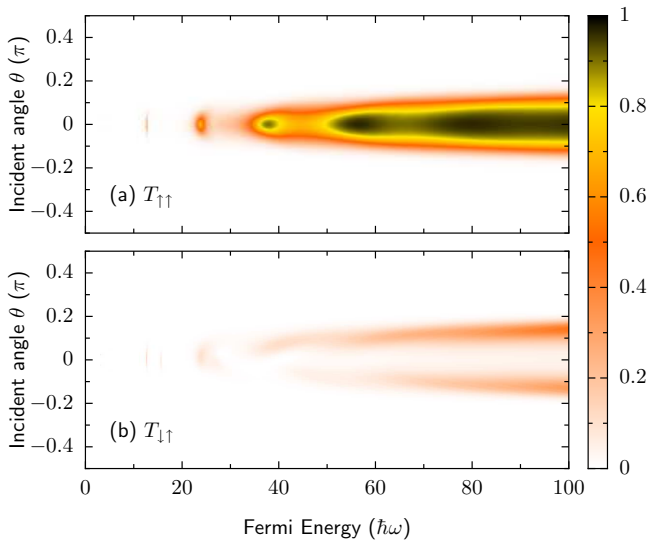}
\caption{(color online) The transmission coefficients (a)
$T_{\uparrow\uparrow}$, (b) $T_{\downarrow\uparrow}$ as function of
energy and incident angle $\theta$ for electrons tunneling through
the magnetic barrier with perpendicular magnetization, $d=2 L_{0}$
and SOI coupling $\alpha_{z}=0.35\times 10^{-11}$ eV$\cdot$m.}
\label{color}
\end{figure}

To evaluate the electron spin polarization effect in the tunneling process,
 we define the net conductance polarization as
\begin{eqnarray}
P=\frac{G_{\uparrow\uparrow}+G_{\uparrow\downarrow}-G_{\downarrow\uparrow}-G_{\downarrow\downarrow}}{G_{\uparrow\uparrow}+G_{\uparrow\downarrow}+G_{\downarrow\uparrow}+G_{\downarrow\downarrow}} \, .
\end{eqnarray}
In Fig.\ \ref{color} we show typical transmission coefficients, as
function of incident energy and angle ($q=\sqrt{2E} \sin \theta$),
for an incoming electron with spin $\uparrow$. The upper panel shows
the transmission coefficient for collected electrons with the same
polarization, $T_{\uparrow\uparrow}$, while the lower is for
outgoing electrons with opposite spins, $T_{\downarrow\uparrow}$.
Here we use a typical SOI strength, $\alpha_{z}=0.35\times 10^{-11}$
eV$\cdot$m, \cite{8} which corresponds to $\alpha_{z}^*=0.5$ .
Notice that for low incident energy and/or with a large incident
angle $\theta$ the transmission coefficients are very small, as the
effective barrier presented by the magnetic strip is quite large
(see bottom panel of Fig.\ \ref{fig1}). At lower incident angle and
higher energy, the transmission exhibits a series of resonances near
normal incidence ($\theta \approx 0$), as the magnetic vector
potential results in an effective scattering double barrier
potential ($V_e \simeq A^2(x)$ in Eq.\  \ref{Ve}) with concomitant
resonances. This structure is also seen in the spin mixing results,
$T_{\downarrow \uparrow}$, as well as in the corresponding spin
$\downarrow$ incidence, $T_{\downarrow \downarrow}$ and $T_{\uparrow
\downarrow}$ (not shown), although with some subtle differences due
to the asymmetry introduced by the field.

\begin{figure}[tp]
\includegraphics[width=1\linewidth]{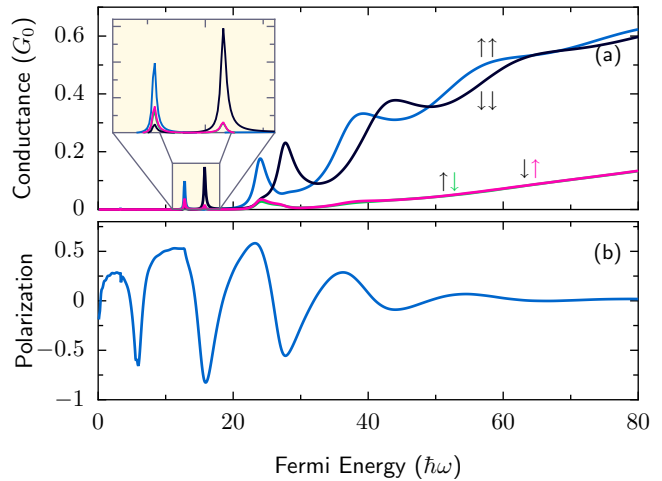}
\caption{(color online) (a) Spin-dependent conductance (in units of $G_{0}$) and (b)
polarization as function of Fermi energy (in units of $\hbar \omega$).  Structure parameters as in Fig.\ \protect\ref{color}.} \label{Graph3}
\end{figure}

The calculated conductance and polarization components for the same parameters as in
 Fig.\ \ref{color} are shown in Fig.\ \ref{Graph3}, {\em vs}.\ Fermi energy.
  The SOI naturally causes the spin of the electron to precess when the electron propagates
  through the gated region.  This is evident in Fig.\ \ref{Graph3}(a)
  as the spin up conductance $G_{\uparrow\uparrow}$ is quite different from
  he spin down conductance, $G_{\downarrow\downarrow}$.
 The inhomogeneous magnetic field also contributes significantly to change
 the tunneling probability of the electron through the magnetic gate barrier.
   $G_{\uparrow\uparrow}$ exhibits peaks shifted in general towards lower energy values,
   in contrast with those in $G_{\downarrow\downarrow}$, a behavior persisting at low energies,
   where the two curves show clearly split resonances.
 In addition, the spin {\em mixing} probability of tunneling electrons depends
 greatly on the incident spin for a given magnetic barrier, as illustrated by
 the curve for $G_{\uparrow\downarrow}$ being slightly different to $G_{\downarrow\uparrow}$
  (see especially the insert in Fig.\ \ref{Graph3}(a)).

The conductance polarization plotted in Fig.\ \ref{Graph3}(b) shows that this
structure results in polarization over 50\% in the first resonant energy and close
to 70\% at the second and higher resonances. Therefore, for realistic values of
 strip magnetization and spin-obit coupling
 the system can generate substantial spin polarized currents,
 {\em even as the injection is unpolarized}.
 Notice that the resonant peaks created by the effective dynamic potential are
 responsible for the large polarization values, as the latter decreases rapidly for higher Fermi energies.

\begin{figure}[tp]
\includegraphics[width=1\linewidth]{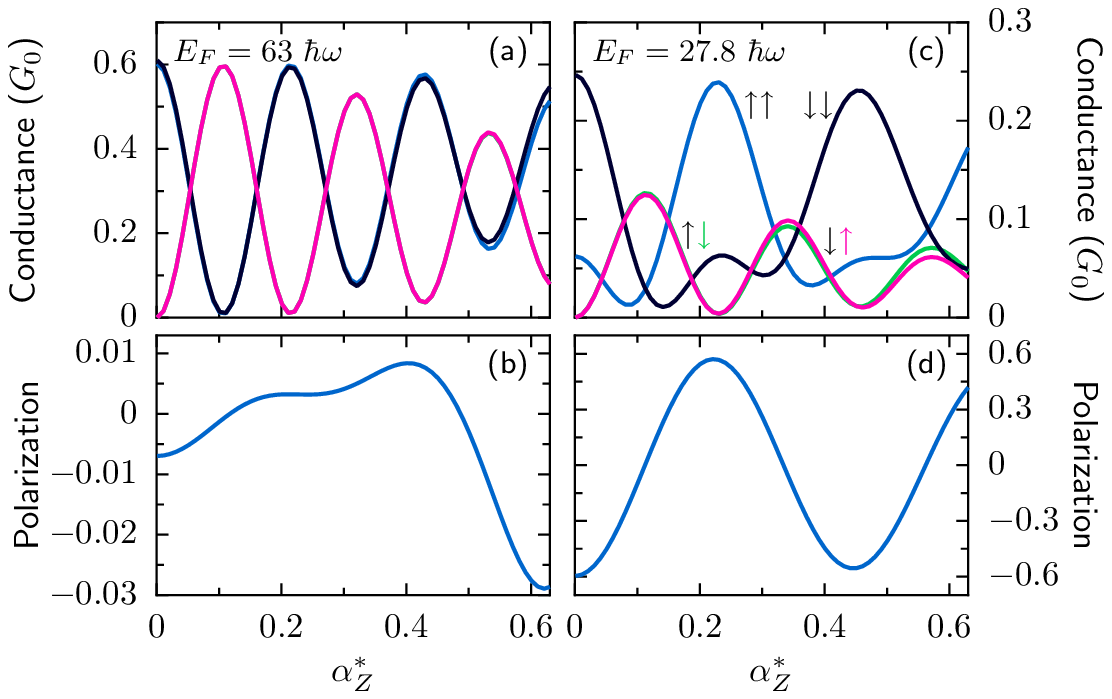}
\caption{(color online) Spin conductance and polarization as function of
$\alpha_z^{*}$ for different Fermi energies.  Notice polarization exceeds 50\% in
panel (d) and reverses in a small $\alpha^*_z$ interval. }
\label{Graph9}
\end{figure}

To provide a more complete account of the role of SOI in this polarization effect we show
in Fig.\ \ref{Graph9} the calculated conductance and polarization as
functions of the SOI strength $\alpha^{*}_{z}$ at two different Fermi
energies. [Notice that $\alpha$ can be varied in experiments by the application
of a gate voltage applied to the metallic strip itself.] Figure \ref{Graph9} (a)
 and (b) show results for $E_{F}=63 \hbar\omega$. At this relatively large energy the spin dependent
conductances show regular oscillations with $\alpha^*_z$, similar to those
expected from the Datta-Das device. \cite{DD} In this region of SOI the
polarization starts near zero (due to Zeeman effect) and has a local maximum at
$\alpha^{*}_{z}=0.45$ of no larger than 1\%, becoming slightly negative afterwards. Figure \ref{Graph9}
(c) and (d) show results for $E_{F}=27.8 \hbar\omega$, near one of the transmission resonances
in $G_{\downarrow \downarrow}$ in Fig.\ \ref{Graph3}.
 It can be seen that increasing $\alpha$ results also in the spin
 conductances having oscillations.  These features reflect the spin precession effects induced by the
SOI, further enhanced by the resonances of the dynamical potential,
as discussed above.  Notice that in this case, the conductance polarization can however
  exceed 50\% and be fully reversed when $\alpha^*_z$ changes by $\simeq 0.23$.

\begin{figure}[tp]
\includegraphics[width=0.8\linewidth]{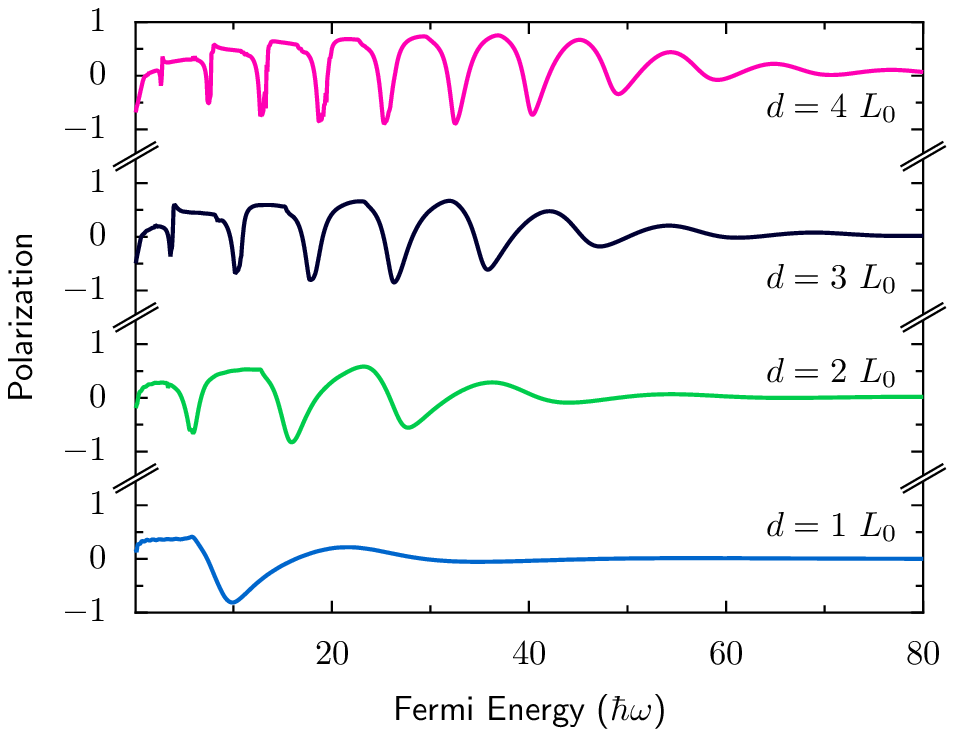}
\caption{(color online) Spin polarization as function of Fermi energy at a given
value of $\alpha_z^*=0.5$ for several width values, $d$, of the magnetic strip. Curves offset for clarity.}
\label{fig5_vb}
\end{figure}

As the strong spin polarization of the conductance is associated with the presence of
transmission resonances through this combination of magnetic and electric field barrier,
we also explore its dependence on strip width--which determines the width of the effective double barrier potential $V_e$.  In Fig.\ \ref{fig5_vb} we show the spin polarization {\em vs}.\ Fermi energy for different magnitudes of the strip width $d$ ($d=L_0$ to $4 L_{0}$). One can see that increasing $d$ results in a larger width of the double barrier potential
well produced by the effective potential, which in turn results in more resonant
peaks in the conductance as the Fermi energy increases.  In all cases, notice that
the polarization exceeds 50\% to 70\%.

We have shown that it is possible to achieve a substantial spin polarization in a realistic situation
 where a magnetized strip is placed in close proximity to a 2DEG.
 Our calculation incorporates the role of the giant Zeeman effect in
  diluted magnetic semiconductors, and this, together with the tunable SOI
  coupling, can be appropriately used to enhance or suppress the polarization.
   We should also notice that as the temperature changes, the effective field can be modulated,
    with the consequent changes on the spin polarization characteristics of a given structure.

The authors thank the support of CMSS and BNNT programs at Ohio University,
as well as NSF grants 0336431 and 0710581. JVB acknowledges support of the  A. von Humboldt Foundation.
 We thank helpful conversations with D. Csontos and M. Zarea.

\end{document}